\documentclass[twoside]{article}
\usepackage{fleqn,espcrc2}
\usepackage{graphicx}
\usepackage{latexsym}
\usepackage{euscript}

\newcommand{\ifig}[1]{\includegraphics[height=80mm,width=75mm]{#1}}
\newcommand{\bc}{\begin{center}}
\newcommand{\ec}{\end{center}}
\newcommand{\be}{\begin{equation}}
\newcommand{\ee}{\end{equation}}
\newcommand{\bsp}{\begin{split}}
\newcommand{\esp}{\end{split}}
\newcommand{\bea}{\begin{eqnarray}}
\newcommand{\eea}{\end{eqnarray}}
\newcommand{\ba}{\begin{eqnarray}}
\newcommand{\ea}{\end{eqnarray}}
\newcommand{\bas}{\begin{eqnarray*}}
\newcommand{\eas}{\end{eqnarray*}}

\newcommand{\simge}{\ \lower-
1.2pt\vbox{\hbox{\rlap{$>$}\lower5pt
\vbox{\hbox{$\sim$}}}}\ }

\newcommand{\AC} {{\cal{A}}}

\newcommand{\AmS}{{\protect\the\textfont2
  A\kern-.1667em\lower.5ex\hbox{M}\kern-.125emS}}
\title{
How To Fix Non-Perturbatively A Parameter Dependent Covariant Gauge On
The Lattice}
\author{L. Giusti\thanks
{This research was supported in part under DOE grant DE-FG02-91ER40676.}
\address{Boston University - Department of Physics,
590 Commonwealth Avenue, Boston MA 02215 USA},
 M. L. Paciello\address{INFN, Sezione di Roma 1,
 P.le A. Moro 2, I-00185 Roma, Italy},
 S. Petrarca\thanks{Speaker at the Conference}$^{\rm b~}$\address{Dipartimento di Fisica, Universit\`a di Roma "La
                     Sapienza"}%
        ,
B. Taglienti$^{\rm b}$
                     }
\begin{document}

\begin{abstract}
We describe how to overcome some problems that usually prevent
from obtaining an efficient  algorithm to fix a generic  covariant  gauge on the lattice.
This gauge is the lattice equivalent of the generic gauge usually adopted in perturbative 
calculations. It depends on a parameter whose value can be varied in order to check
the gauge dependence of measured  matrix elements.
\vspace{1pc}
\end{abstract}

\maketitle

We discuss how to enforce on the lattice the 
generic covariant  gauge:
\be
{\partial_{\mu}}{A_{\mu}}^{G(\alpha)}  (x)=\Lambda(x) \; ; 
\label{eq:dinamic1}
\ee
where $\Lambda(x)$  are  matrices belonging to the Lie algebra of the
 SU(3) group and $ \alpha$ is the gauge parameter  which is the width 
of the gaussian distribution used to extract the $\Lambda(x)$ matrices  as will be described
in the following. The differential equation (\ref{eq:dinamic1}) is supplied
with periodic boundary conditions
on the lattice.

\begin{figure}[ht]
\bc
\ifig{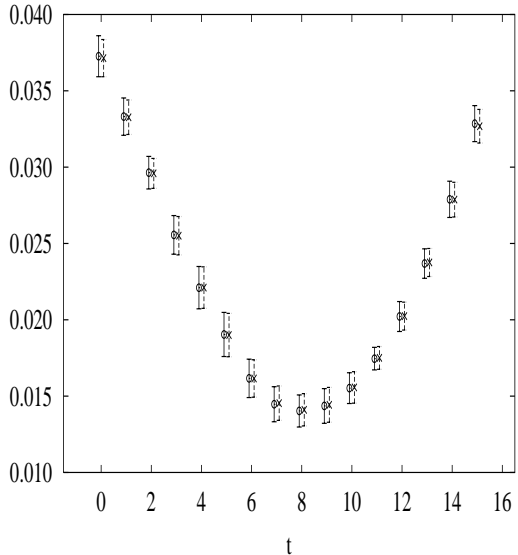}
\caption{\small{Comparison of the matrix elements of  $\langle\AC^{'}_i\AC^{'}_i\rangle(t)$ 
(crosses) and the rescaled $\langle\AC_i\AC_i\rangle~\cdot~C_i^2(g_0)$
(open circles) as function of time for 
a set of 50 thermalized $SU(3)$  configurations at $\beta=6.0$ with a 
volume $V\cdot T=8^3\cdot 16$, $C_i =.729(2)$. The data have been
slightly displaced in $t$ to help eye, the errors are jacknife. This
figure is taken from ref. \cite{giusto}.}}
\label{fig:amu}
\ec
\end{figure}

Before going into details, we want to make a 
preliminary remark concerning the definition  of the
gauge potential $A_\mu$ on the lattice. As it is well known the gauge potential,
on the lattice, is a derived  quantity   being the links $U_\mu$ the natural
fields. The link is the parallel transporter  of the theory and is connected
with the gauge potential $A_\mu$ only in the continuum limit.
Of course, it is necessary to choose a definition of $A_\mu$
in terms of  the links in order to fix the gauge; this is not a problem because any $A_\mu$ definition  
differing from another  by irrelevant terms ( $O(a)$) 
must go to the same continuum operator up
to a constant. This trivial, though fundamental, consideration was checked 
non-perturbatively by numerical simulations,
\cite{giusto} obtaining a spectacular coincidence (see Fig.~\ref{fig:amu}) between the rescaled
values of the correlator:
\be
\langle \AC_i\AC_i\rangle (t) \equiv \frac{1}{3 V^2}  
 \sum_{i}\sum_{{\bf x},{\bf y}} Tr \langle  A_i({\bf x},t)A_i({\bf y},0)\rangle \; ,
\label{eq:AiAi}\\
\ee
built up assuming the following different $A_\mu$ definitions:

 \be
A_{\mu} (x) \ 
\equiv \ {{( U_{\mu} (x) - U_{\mu}^{\dagger} (x) )_{traceless}}\over
{2 i a g_0}}\; ,
\label{eq:prima}
\ee
\be
{A^{'}}_{\mu} (x) \
\equiv \ {{( (U_{\mu} (x))^2 - (U_{\mu}^{\dagger} (x))^2 )_{traceless}}
\over {4 i a g_0}}\; .
\label{eq:seconda}
\ee

We found that the expected  relation  $A'_\mu (x) = C(g_0) A_\mu (x)+ O(a^2)$
is fulfilled and the effect  of order $a$ terms is negligible in the measured matrix elements.
 This result substantially supports
the correctness of the continuum limit of matrix elements built up by the insertion of 
operators representing $A_\mu$  on the lattice.

Actually the freedom of the $A_\mu$ definition on the lattice, instead of being considered
 problematic,
has been rarely exploited  in literature and it is 
interesting to quote two recent papers that take into account this possibility for different
purposes. Ref. \cite{furui} proposed to use, in the Landau gauge, the 
exponential form for the relation between $U_\mu$ and $A_\mu$  instead of the
linear one (\ref{eq:prima}).
In ref. \cite{williams} it has been attempted  to ``improve'' the Landau gauge 
fixing by studying different  $A_\mu$ definitions.

The classical procedure  to fix a generic covariant gauge on the lattice
can be read from the  following  expression of the expectation value of a gauge
dependent operator ${\cal O}$ on the lattice:
\bea
\langle{\cal O}\rangle&=&\frac{1}{Z}\int d\Lambda \int dU  \nonumber \\
& &{\cal O}({U^{G(\alpha)}})
e^{-\frac{1}{2 \alpha}\int d^4x 
Tr(\Lambda^2)}
e^{-\beta S(U)} \; ;
\label{eq:omedio}
\eea
where $S(U)$ is the Wilson lattice gauge invariant action, 
and $G{(\alpha )}$ is the gauge transformation that enforces
the gauge condition, eq. (\ref{eq:dinamic1}).

Following the usual procedure,
the gauge transformation $G(x)$ must be determined numerically by
applying a steepest descent iterative algorithm which minimizes an ad hoc functional 
$F[G]$ chosen in order to have minimum (extremum)  corresponding to a gauge
transformation  satisfying the gauge condition.

The definition of $<{\cal O}>$ shows that before applying to each thermalized
configuration  the gauge fixing algorithm one has to generate a set of $\Lambda$,
one for each lattice site, distributed as a gaussian of width~$\alpha$.
Instead of taking more different sets of $\Lambda$ as it is
suggested by (\ref{eq:omedio}) and then averaging, it is reasonable
to take only one set of $\Lambda$ for each configuration. 
In fact one generates 
for each thermalized configuration a different  sets of $\Lambda$ matrices
with the same width $\alpha$.

The  functional  $F[G]$, that in the Landau gauge is given by
$
F_L[G]\equiv ||A^G||^2=
\int\mbox{\rm Tr}\left(A^G_{\mu}(x)A^G_{\mu}(x)\right)d^4x
$, 
must be changed in order to reach the new gauge condition.
This problem was discussed in \cite{giusti} where the following more complicated  functional form
was suggested:
\be
F[G]=\int d^4x\mbox{\rm Tr}\left[(\partial_{\mu}A^{G(\alpha)}_{\mu}-\Lambda)^2  
\right]\; .  
\label{eq:cov11}
\ee
The stationary points of this functional  correspond to the following
gauge condition:
\be
D_{\nu}\partial_{\nu}(\partial_{\mu}A^{G(\alpha)}_{\mu}-\Lambda)=0\; .
\label{eq:casino}
\ee
Of course, when 
$\partial_{\mu}A^{G(\alpha)}_{\mu}(x)-\Lambda(x)=0$, 
the new functional is stationary 
but there can be other non trivial solutions.
This fact is obviously a problem of the new functional, 
and simpler functional forms are difficult to find, as it has been discussed
in \cite{giusti}. We will assume the functional
form (\ref{eq:cov11}) and we will ignore the problem of spurious solutions in the following.
The choice of the form of $F[G]$ is not sufficient to solve the numerical
problem of fixing the generic covariant gauge because, due to the 
complicated  structure of (\ref{eq:casino}), it is likely that the usual gauge
fixing algorithm  either does not converge or takes too much computer time.
This problematic convergence  can be traced back to the following
reason. In the Landau case, being $F_L$ proportional to $U_\mu$, when the gauge
fixing algorithm visits the lattice point $ {\tilde x} $ it puts only the gauge
transformation in the point $ {\tilde x} $ different to the identity and then 
locally $F_L$ depends linearly on $G( {\tilde x})$. Of course this property is
not valid anymore for a na\"{\i}ve transcription on the lattice of $F[G]$
because this shows a natural quadratic dependence on $G( {\tilde x})$.

At this point one can take advantage of the freedom to choose the discretisation 
of $A_\mu$.  In fact it is possible to find a discretisation 
(``driven discretisation'') for  the functional $F[G]$
such that it takes only a local linear dependence on $G ( {\tilde x})$.
This aim can be
reached by allowing different terms of $F$ to be chosen in such a way to guarantee the
local linear dependence on $G({\tilde x})$, instead of being an algebraic consequence of a
particular definition of $A_\mu$.
In the following we show some examples of the discretisation for the
terms contained  in the functional $F[G]$:
\bas
\lefteqn{2iag A_\mu(x) = U_\mu(x) - U_\mu^\dagger (x);}\\
\lefteqn{a^2 g^2 A_\mu(x) A_\nu(x) = } \\
\lefteqn{\quad I + U_\mu^\dagger(x) U_\nu(x) - U_\nu(x) 
                                - U_\mu^\dagger(x);}\\
\lefteqn{ a^2 g^2 A_\mu(x-\mu) A_\nu(x-\nu) =I+}\\ 
\lefteqn{\quad U_\mu(x-\mu) U_\nu^\dagger(x-\nu) - U_\mu(x-\mu) 
                      - U_\nu^\dagger(x-\nu).}
\eas
Each term locally transforms linearly in $G({\tilde
x})$ and it is 
a possible discretisation of the continuum operator written on the left side
of each expression.
Using the driven discretisation the expression of $F[G]$
assumes the following  compact form:
\bas
F[G]  =  \frac{1}{V a^4 g^2} Tr \sum_{x} J^G(x) 
J^{G \dagger }(x) \quad ,
\eas
where 
\bas
J(x)  =  -8 I + \sum_{\nu} \left[ U_\nu^\dagger(x-\nu) + U_\nu(x) \right]
 -i a g \Lambda \quad.
\eas
\begin{figure}[htb]
\bc
\ifig{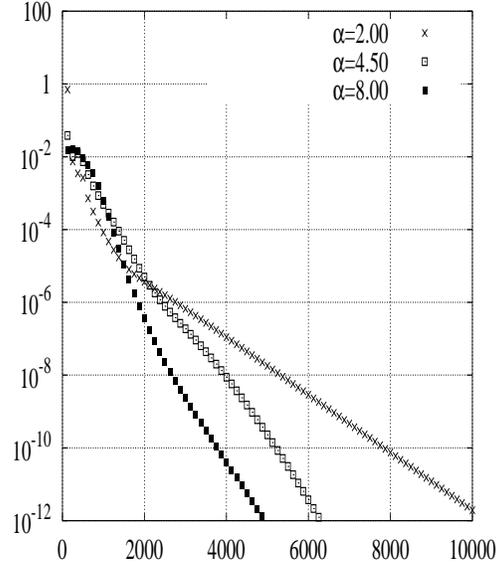}
\caption{\small{Behaviour of $\delta F$ vs gauge fixing sweeps (with
overrelaxation) for different
values of $\alpha$; \mbox{$\beta=5.8$}, volume=$6^4$. }}
\label{fig:discesa}
\ec
\end{figure}
As it is shown in Fig.~\ref{fig:discesa}, the number of sweeps necessary in order
to reach a prefixed quality of the gauge fixing increases when $\alpha$ decreases.
Each sweep takes around $ 20 \%$ more computer time than in the case of the functional
$F_L$. The quality of the gauge fixing is measured by adopting as quality factor
the variation $\delta F$ of $F[G]$ between a sweep and the next.

Further details on the formalism, spurious solutions and checks will be discussed
in a forthcoming paper (Ref. \cite{pape}).

\end{document}